\documentclass{aa}
\usepackage[dvips]{graphicx}
\usepackage{psfig}
\begin{document}
 \title{Discovery of a widely separated binary system \\
        of very low mass stars
   \thanks{Based on observations made at the European Southern 
	Observatory, La Silla, Chile}}
   \author{N.~Phan-Bao
       \inst{1}
       \and
       E.L.~Mart\'{\i}n
       \inst{2,3}
       \and
       C.~Reyl\'e
       \inst{4}
       \and
       T.~Forveille
       \inst{5,6}
       \and
       J.~Lim
        \inst{1}
       }

\offprints{N.~Phan-Bao, \email{pbngoc@asiaa.sinica.edu.tw}}

\institute{Institute of Astronomy and Astrophysics, Academia Sinica.
           P.O. Box 23-141, Taipei 106, Taiwan, R.O.C.
          \and
	  Instituto de Astrof\'{\i}sica de Canarias, C/ V\'{\i}a L\'actea  
          s/n, E-38200 La Laguna (Tenerife), Spain.
          \and
          University of Central Florida, Dept. of Physics, PO Box 162385, 
          Orlando, FL 32816-2385, USA.
          \and
	  CNRS UMR6091, Observatoire de Besan\c{c}on, BP1615, 
	    25010 Besan\c{c}on Cedex, France.
	  \and
	  Canada-France-Hawaii Telescope Corporation, 65-1238 Mamalahoa 
          Highway, Kamuela, HI 96743 USA.	
	  \and
	  Laboratoire d'Astrophysique de Grenoble, Universit\'e J. 
          Fourier, B.P. 53, F-38041 Grenoble, France.
 }

      \date{Received / Accepted}

\abstract{We report our discovery of a nearby wide binary system, 
LP~714-37~AB. Phan-Bao et al. identified LP~714-37 as a mid-M dwarf during a 
cross-identification of the NLTT and DENIS catalogues. Our CCD images 
resolve the system into a binary with a projected angular separation 
of 1.8\arcsec, or 33~AU, and low-resolution optical spectra give
spectral types of M5.5 (LP~714-37A) and M7.5 (LP~714-37B). This makes
LP~714-37~AB one of very few widely separated (separation $>$ 30 AU) 
very low-mass binary systems known in the field, and we discuss it 
in the context of the multiplicity properties of very low-mass stars and 
brown dwarfs.  
 }           
\titlerunning{A widely separated binary system}
\authorrunning{N.~Phan-Bao et al.}
  \maketitle


\section{Introduction}
Binary systems have been studied for decades to measure accurate stellar
masses, and to test evolutionary models and star formation theories.
Considerable attention has recently been paid to very low-mass (VLM) 
companions to low-mass stars (Duquennoy \& Mayor \cite{duquennoy91}; 
Fischer \& Marcy \cite{fischer}; Delfosse et al. \cite{delfosse99};
Reid et al. \cite{reid01}; Beuzit et al. \cite{beuzit04}; Forveille
et al. \cite{forveille04}), as well as to binaries among ultracool 
dwarfs (spectral types later than M6) in the solar neighborhood 
(Mart\'\i n et al. \cite{martin99a}; Close et al. \cite{close02}, 
\cite{close03}; Bouy et al. \cite{bouy}; Burgasser et al. \cite{burgasser03}; 
Gizis et al. \cite{gizis03}; Siegler et al. \cite{siegler}; Forveille
et al. \cite{forveille05}) and in nearby young open clusters and associations
(Mart\'\i n et al. \cite{martin98}, \cite{martin00}, \cite{martin03};
Chauvin et al. \cite{chauvin}).
The main results of the high spatial resolution imaging surveys of
VLM stars and brown dwarfs (BDs) can be summarized as follows: (1) the 
binary frequency in the separation range 1-15 AU is about 15\% , (2) 
the frequency of wide binary systems (semi-major axis $>$ 15 AU) is
very low, $<$1\% , (3) the mass ratios are strongly biased towards 
nearly-equal mass binaries, beyond the expected observation selection
effects. The companions to the more massive low mass stars, by contrast,
span a wider range of both separations and mass ratios.

The properties of VLM binaries are an important constraint for models 
of star-formation and evolution. It has been debated in the literature 
whether the properties of VLM binaries and stellar binaries differ, 
implying different formation mechanisms (Kroupa et al. \cite{kroupa03}),
or whether the binary properties instead show continuous trends with 
decreasing primary mass, implying that VLM binaries form through the same
processes as stellar binaries (Luhman \cite{luhman04b}). Clearly there is 
a need for a larger sample of observed VLM binaries, particularly at wide
separations where few of them are known. One leading model of BD formation
is that they form and are ejected in unstable multiple systems within
small clusters (Bate et al. \cite{bate02}, \cite{bate03}). Numerical  
simulations of decaying N-body clusters indicate that the typical separations
of binaries composed of stars with masses ranging between 0.1~$M_{\odot}$ 
and 0.2~$M_{\odot}$ would be below 10~AU (Sterzik \& Durisen \cite{sterzik}),
in rough agreement with the observations. The ejection models (Reipurth 
\& Clarke \cite{reipurth}; Bate et al. \cite{bate02}) suggest that the
binary BD systems that do exist must be close (separations~$\leq$~10~AU).
The detection of wide VLM binary systems has thus become an important 
test of the ejection models. The first wide binary BDs have been found in 
young ($<$10~Myr) associations or clusters: 2MASS~J1101-7732 
(240~AU separation, in Chamaeleon~I, Luhman \cite{luhman04a}); 
2MASS~J1207-3932 (55~AU, in TW Hydrae, Chauvin et al. \cite{chauvin}).
Their existence is at first sight inconsistent with the ejection models,
though the statistics of the numerical simulations is currently limited.

In this Letter, we present a new wide binary consisting of two VLM stars
in the field. Sec.~2 presents the observations and data reduction, while
Sec.~3. discusses the results in the context of the binary properties of 
VLM stars and BDs. 
\section{Observations and data reduction}
Phan-Bao et al. (\cite{phan-bao03}) identified LP~714-37 (DENIS-P~J0410-1251) 
as a mid-M dwarf while cross-identifying the NLTT (Luyten \cite{luytenb}) 
and DENIS (Epchtein \cite{epchtein97}) catalogues, and derived
a photometric distance of 15.3~pc. Cruz et al. (\cite{cruz03}) 
independently determined a spectrophotometric distance of 15.4~pc and
an M6.0 spectral type. Allowing for the binary nature of the system
pushes its distance out to 18.1~pc (Sec.~3).

\begin{figure}[t]
\hspace{1.2cm}
\psfig{width=6.0cm,file=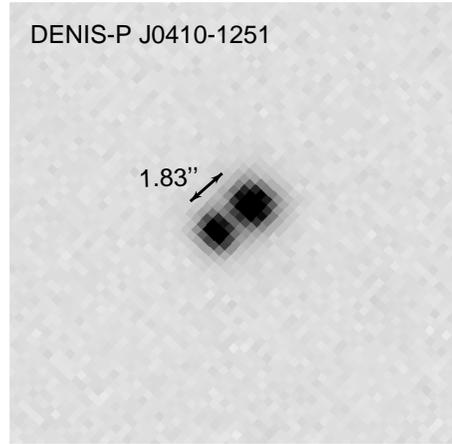,angle=0}
\caption{A 20$\times$20\arcsec section of the R-band NTT image of the 
LP~714-37 binary system. The pixel size is 0.33\arcsec, North is up 
and East is left.}
\label{fig_image}
\end{figure}

As part of our ongoing spectroscopic follow-up of new nearby M dwarfs 
detected with DENIS, one of us (C\'eline Reyl\'e) observed LP~714-37 
with the 3.6-meter New Technology Telescope (NTT) at La
Silla Observatory (ESO, Chile) in November 2003;
and the data was analyzed in January 2005.
Two objects were in fact detected
on the acquisition image 
(Fig.~\ref{fig_image}).

\begin{figure}[t]
\hspace{0.7cm}
\psfig{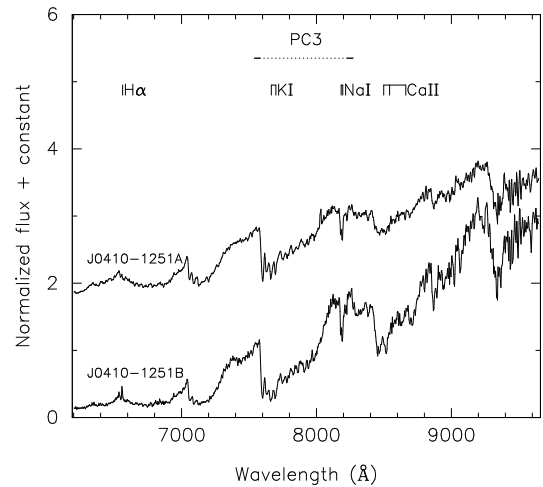}
\caption{Spectra of the two components of LP~714-37. The positions of 
the H$_{\alpha}$, NaI, KI and CaII lines are indicated, as well as the 
spectral intervals used to compute the PC3 spectral index.}
\label{fig_spectre}
\end{figure}

Optical low-resolution spectra were obtained for both objects in the Red 
Imaging and Low-dispersion spectroscopy (RILD) observing mode of the EMMI 
instrument. The detector was a 2048$\times$4096 MIT/LL CCD, used in normal 
readout mode with 2$\times$2 binning. Grism\#2 gave a dipersion of 
3.5~\AA~pixel$^{\rm -1}$, and the OG530\#645 order sorting filter
defined a wavelength range of 520 to 950 nm. The slit width of 1 arsec 
corresponds to a spectral resolution of 10.4~\AA, and the seeing was 
0.7 arcsec. The exposure time was 30~s. LTT~2415 and Feige~110 (selected 
from the ESO list, ftp.eso.org/pub/stecf/standards/ctiostan/) were 
observed as spectrophotometric standards.
The reduction of spectra was performed with MIDAS packages.
We normalized the spectra to 1 over the 754-758~nm interval, 
the denominator of the PC3 index (Mart\'{\i}n et al. \cite{martin99b}) 
and a region with a good flat pseudo-continuum. Figure~\ref{fig_spectre} 
shows the two resulting spectra. The presence of the Na{\small I} 
and K{\small I}  doublets and the absence of the Ca{\small II} triplet 
immediately demonstrate that both stars are M dwarfs rather than
M giants, as does the general shape of the TiO bands.
Analysis of the R band acquisition image with SExtractor (Bertin \& Arnouts 
\cite{bertin}) gives a separation of 5.53 pixels, or 1.83\arcsec~with
the 0.33\arcsec/pixel focal scale of EMMI. 
Since the discovery is serendipitous we did not obtain images
with other filters and have no photometric calibrations for the R
band. We therefore only use the image for astrometry.
%
\begin{figure}[t]
\hspace{0.7cm}
\psfig{width=7.0cm,file=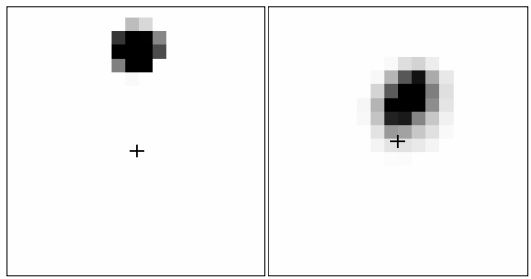,angle=0}
\caption{Archival images of LP 714-37: SERC-I (left, epoch: 1986.940) and
DENIS-I (right, epoch: 2000.896). The cross indicates the position of 
component B at the 2003.915 epoch of the NTT image. Clearly, the SERC-I 
image would easily separate the two objects if they were not physically 
associated. The size of each image is 20$\times$20\arcsec, and North is 
up and East to the left.}
\label{fig_pm}
\end{figure}
\begin{table*}
   \caption{PC3 indices and spectral types of the two components of LP~714-37}
    \label{table_indices}
  $$
   \begin{tabular}{lllllllllllll}
   \hline 
   \hline
   \noalign{\smallskip}
Stars               &Other name & PC3  & SpT  &$M_{\rm I}$& $M_{\rm J}$ & $M_{\rm H}$ & $M_{\rm K}$& Mass  &  Sep.       & P   &  dist.  & $\mu$   \\
                    &           &      &      &       &         &         &        &(M$_{\odot}$)& (AU)  &  (yr) &  (pc) & \arcsec/yr   \\           
  (1)               &(2)        &(3)   &(4)   &(5)    &(6)      & (7)     &    (8) &  (9)   & (10)       & (11) & (12)   & (13)    \\
    \noalign{\smallskip}
    \hline 
LP 714-37A & DENIS-P J0410-1251 & 1.32 & M5.5 & 11.96 &~\,9.95  &~\,9.34  &   9.11 &  0.11 &  33.1$\pm$4.0   &  426 &  18.1$\pm$2.2 & 0.400    \\
LP 714-37B &                    & 1.72 & M7.5 & 13.56 & 10.85   &  10.24  &   9.82 &  0.09 &                 &      &                 &     \\
    \noalign{\smallskip}
    \hline 
   \end{tabular}
  $$
  \begin{list}{}{}
  \item[] 
{\it Columns 1} \& {\it 2}: NLTT and DENIS name.
{\it Columns 3} \& {\it 4}: The PC3 index and spectral types derived from the (PC3, spectral type)
relation of Mart\'{\i}n et al. (\cite{martin99b}).
{\it Columns 5-8}: Absolute magnitudes for the $I$, $J$, $H$, and $K$ bands based on the PC3-absolute magnitudes relation
(Crifo et al. \cite{crifo}). 
{\it Column 9}: Mass determinations for 1-5~Gyr from the models of Baraffe et al. (\cite{baraffe98}).
{\it Column 10}: Projected separation.
{\it Column 11}: Period computed by assuming a face-on circular orbit.
{\it Columns 12}: Spectrophotometric distance.
{\it Columns 13}: Total proper motion taken from Phan-Bao et al. (\cite{phan-bao03}).
  \end{list}
\end{table*}
%

\section{Discussion}
The calibration of the PC3 index 
to spectral type (Mart\'{\i}n et al. \cite{martin99b}) gives spectral types of M5.5 for component A and M7.5 
for component B, with an uncertainty of $\pm$0.5 subclass. The observed 
H$_{\alpha}$ emission in component B with spectral type of M7.5 is 
consistent with the observations by Gizis et al. (\cite{gizis00a}) 
pointed out that all M7-M8 dwarfs show the H$_{\rm \alpha}$ line
in emission.

The proper motion of the system is 
$\mu_{\rm \alpha}$ = $-$117~mas/yr and $\mu_{\rm \delta}$ = $-$382~mas/yr
(Phan-Bao et al. \cite{phan-bao03}), and it has therefore moved by
6.8~arcsec between the epoch of the SERC-I plate and our NTT observation.
Figure~\ref{fig_pm} shows that there is no background star at the position
of the system in either the SERC-I image ($I$ band) or the DENIS-I image,
and therefore demonstrates that the system is a physical binary.
\begin{figure}[t]
\hspace{0.7cm}
\psfig{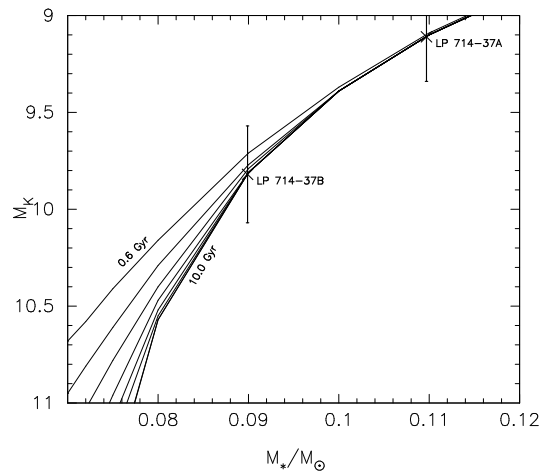}
\caption{The mass vs. K band absolute magnitude relations from the
evolutionary models of Baraffe et al. (\cite{baraffe98}) for solar metallicity
low mass stars. The crosses indicate the masses of components A and B 
for an age of 5~Gyr.
The isochrones plotted are 0.6, 0.8, 1.0, 1.3, 1.6, 2.0, 5.0, and 10.0 Gyr; 
the latter two are overlapped.}
\label{B98}
\end{figure}
\begin{figure}[t]
\hspace{0.4cm}
\psfig{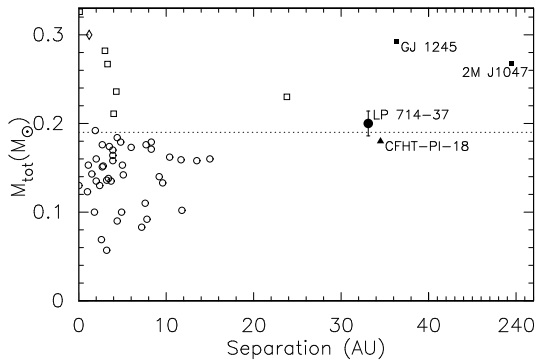}
\caption{Sum of primary and secondary component masses vs. projected 
binary separation for VLM binaries in the field. The empty circles 
represent binaries with separations smaller than 30~AU, from Table~4 of 
Siegler et al. (\cite{siegler}) and references therein. 
The empty squares show low mass field binaries from Reid \& Gizis 
(\cite{reid97a}) with masses updated from RECONS at 
http://www.chara.gsu.edu/RECONS/. 
The empty rhombus is LHS~224 from Beuzit et al. (\cite{beuzit04}).
The 35~AU VLM binary CFHT-Pl-18 (Mart\'{\i}n et al. 
\cite{martin00}) is plotted as a filled triangle.
The filled squares show two VLM triple systems GJ~1245 and 2MASS 
J1047+4026, discussed in Sec.~3. The filled circle indicates 
LP~714-37. Dotted line: M$_{\rm tot}=0.185$~M$_{\odot}$, the
limit for VLM systems adopted by Close et al. (\cite{close03}).   
}
\label{separation}
\end{figure}
%

%
%

To estimate the distance, we assume that this system has no additional
unresolved component(s). We can then calculate the absolute magnitudes 
of the system from those of the two components, computed from their PC3
indices with the PC3 index to magnitudes relation of Crifo et al. 
(\cite{crifo}) (Table~\ref{table_indices}). Comparison with 
the DENIS apparent magnitudes ($I=12.99$, $J=10.94$, $K=9.89$)
then gives the distance: $d_{\rm I}=17.8$$\pm 2.1$~pc; 
$d_{\rm J}=18.9$$\pm 2.3$~pc
 and $d_{\rm K}=17.6$$\pm 2.1$~pc. 
We adopt the mean distance of $d=18.1$$\pm 2.2$~pc.

We estimate the mass of each component, using $I$, $J$, and $K$-band 
mass-luminosity relations (Fig.~\ref{B98} for the $K$ band) from the 
Baraffe et al. (\cite{baraffe98}) models. We adopt their 5~Gyr 
model, an intermediate age for the solar neighbourhood (Caloi et al. 
\cite{caloi}), but the results are insensitive to this choice 
(Fig.~\ref{B98}). The masses of component A and B are respectively 
0.11$\pm$0.01~M$_{\odot}$ and 0.09$\pm$0.005~M$_{\odot}$, slightly above 
the hydrogen-burning limit (Chabrier \& Baraffe \cite{chabrier}).
The total mass of LP~714-37AB is 0.2$\pm 0.015$~M$_{\odot}$, within
one sigma of the $M_{\rm tot}<0.185~M_{\odot}$ convention adopted 
by Close et al. (\cite{close02}) for their sample of VLM binaries. Thus, 
this system can be considered a VLM binary.
At the 18.1~pc of the system, its 1.83\arcsec~separation corresponds to 
33.1~AU. Assuming a face-on circular orbit, the corresponding orbital period 
is approximately 400~years.

Recent surveys demonstrate that VLM binaries with large separations 
($>$~30~AU) are rare in the field (Fig.~\ref{separation}, and 
references therein), but can be found in young associations and 
clusters (Chauvin et al. \cite{chauvin}, Luhman \cite{luhman04a}).
Those authors reported the respective discoveries of 2MASSW~J1207-3932 
(8~Myr, TW Hydrae), and 2MASSW~J1101-7732 (1~Myr, Chamaeleon~I)
with separations of 55~AU and 240~AU. We report here the discovery of 
a 33~AU VLM binary, and Mart\'{\i}n et al. (\cite{martin00}) found that
CFHT-Pl-18 is a 35~AU VLM binary (Fig.~\ref{separation}), both in the 
field.

Since ejection models predict a maximum separation of about 10~AU for 
VLM binaries, the existence of the wide binaries is at first sight 
inconsistent with these ejection models. One should note however that
the numerical models to date suffer from small number statistics. A
further caveat if that the relevant quantity is the total mass of the
system, and that the apparent binaries could possibly be triple or 
higher order multiple systems, with a correspondingly higher total mass.
This would make them analogs of the GJ~1245ABC triple system, which
consists of two M5.5 and one $\sim$M8 dwarfs (Reid et al. \cite{reid95},
Henry et al. \cite{henry99})
with separations of 32 and 5 AUs (McCarthy et al. \cite{mccarthy88}).
We note that Gizis et al. (\cite{gizis00b}) discovered a 230~AU VLM binary 
in the field, LP 213-67 (M6.5) and LP 213-68 (M8.0), which in the recent 
adaptive optics survey of Close et al. (\cite{close03}) turned out to be 
triple when LP 213-67AB (or 2MASS J1047+4026) was resolved into two components (see 
Fig.~\ref{separation}). Bouy et al. (\cite{bouy05}) further reported 
that the DENIS-P~J020529.0-115925 VLM binary is a probable triple system. 
Triple systems could thus potentially explain the apparent excess of
wide VLM binaries, and adaptive optics imaging of LP~714-37 would be
of obvious interest to clarify its true multiplicity.

Amongst the recently discovered VLM field binaries (Bouy et al. 2003; 
Close et al. \cite{close02}, \cite{close03};  Siegler et al. \cite{siegler}; 
Forveille et al. \cite{forveille05}), LP~714-37 has one of the widest
separation, making it of great interest as a constraint for VLM binary 
star formation theories. We note that late-M dwarfs detected by Phan-Bao 
et al. (\cite{phan-bao01}, \cite{phan-bao03}) over 5700 square degrees 
in the DENIS database provide a valuable well defined sample for studies
of VLM field binaries.

\begin{acknowledgements}
Ngoc Phan-Bao is grateful to the DENIS consortium for access to 
the DENIS data used by his very low mass stars search. Ngoc Phan-Bao 
also thanks Fran\c{c}oise Crifo for her comments on the manuscript.
Celine Reyl\'e acknowledges help during the observations by Olivier 
Hainaut and the NTT team at the European Southern Observatory.
We thank the referee, Kevin Luhman, for a fast and useful report.
Partial funding was provided 
by NSF grant AST 02-05862.

\end{acknowledgements}

\end{document}